\documentclass[12pt,preprint]{aastex}

\newcommand{\ii}{$i'$}
\newcommand{\zz}{$z'$}

\newcommand{\bb}{$B$}
\newcommand{\vv}{$V$}
\newcommand{\jj}{$J$}
\newcommand{\hh}{$H$}

\newcommand{\iw}{$I$}
\newcommand{\ts}{\thinspace}

\newcommand{\etal}{{et\thinspace al.}}

\newcommand{\Ang}{\AA\thinspace}

\newcommand{\Ho}{$H_{0}$}

\newcommand{\tabref}[1]{Table~\ref{#1}}
\newcommand{\figref}[1]{Figure~\ref{#1}}
\newcommand{\secref}[1]{\S~\ref{#1}}

\begin{document}

\title{Starburst Intensity Limit of Galaxies at $z\!\simeq\!5\!-\!6$}

\shorttitle{Starburst Intensity Limit at $z\!\simeq\!5\!-\!6$}

\author{N. P. Hathi\altaffilmark{1}, S. Malhotra\altaffilmark{1,2} and J. E. Rhoads\altaffilmark{1,2}}

\altaffiltext{1}{Department of Physics, Arizona State University, Tempe, AZ 85287-1504, USA}

\altaffiltext{2}{School of Earth and Space Exploration, Arizona State University, Tempe, AZ 85287-1404, USA}

\email{Nimish.Hathi@asu.edu}
\shortauthors{Hathi et al}


\begin{abstract}
The peak star formation intensity  in starburst galaxies does not vary
significantly from  the local  universe to redshift  $z\!\sim\!6$.  We
arrive at this conclusion  through new surface brightness measurements
of  47  starburst   galaxies  at  $z\!\simeq\!5\!-\!6$,  doubling  the
redshift   range   for   such   observations.   These   galaxies   are
spectroscopically  confirmed in  the  Hubble Ultra  Deep Field  (HUDF)
through  the  GRism ACS  program  for  Extragalactic Science  (GRAPES)
project.    The   starburst    intensity   limit   for   galaxies   at
$z\!\simeq\!5\!-\!6$  agree  with  those at  $z\!\simeq\!3\!-\!4$  and
$z\!\simeq\!0$  to within  a factor  of  a few,  after correcting  for
cosmological  surface  brightness  dimming  and for  dust.   The  most
natural interpretation of this constancy  over cosmic time is that the
same physical mechanisms limit starburst intensity at all redshifts up
to $z\!\simeq\!6$ (be  they galactic winds, gravitational instability,
or something else).  We do see two trends with redshift: First, the UV
spectral slope ($\beta$) of  galaxies at $z\!\simeq\!5\!-\!6$ is bluer
than that  of $z\!\simeq\!3$ galaxies, suggesting an  increase in dust
content over  time.  Second, the  galaxy sizes from  $z\!\simeq\!3$ to
$z\!\simeq\!6$   scale   approximately   as   the   Hubble   parameter
$H^{-1}(z)$.   Thus,  galaxies  at  $z\!\simeq\!6$ are  high  redshift
starbursts, much  like their local  analogs except for  slightly bluer
colors,  smaller  physical sizes,  and  correspondingly lower  overall
luminosities.   If we  now assume  a constant  maximum  star formation
intensity,  the  differences in  observed  surface brightness  between
$z\!\simeq\!0$  and   $z\!\simeq\!6$  are  consistent   with  standard
expanding cosmology and strongly inconsistent with tired light model.
\end{abstract}



\section{Introduction}\label{introduction}

Star formation on galactic scales is a key ingredient in understanding
galaxy evolution.  We  cannot compare structure formation calculations
to  observed  galaxy  populations  without  some model  for  how  star
formation proceeds.  Such models are based on detailed observations in
the nearby  universe, combined  with physically motivated  scaling for
differing conditions elsewhere in  the universe.  To test the validity
of such scaling, it is  valuable to directly measure the properties of
star  formation events  in  the  distant universe,  and  see how  they
compare with their nearby counterparts.

Starbursts  are regions  of intense  massive star  formation  that can
dominate a galaxy's integrated  spectrum.  By comparing the properties
of starbursts over a wide range  of redshifts, we can test whether the
most  intense  star formation  events  look  the  same throughout  the
observable  history  of  the  universe.  High  redshift  galaxies  are
expected, on average, to be  less massive and lower in metal abundance
than their present-day counterparts.  Either effect could in principle
change the maximum intensity of  star formation that such galaxies can
sustain.
   
\citet{meur97}   (hereafter  M97)   measured  the   effective  surface
brightness, i.e.,  the average  surface brightness within  an aperture
that encompasses half  of the total light, for  various samples.  They
conclude that  the maximum  effective surface brightness  of starburst
galaxies  is unchanged to  better than  an order  of magnitude  out to
redshifts  $z\!\simeq\!3$.   \citet{weed98}  (hereafter W98)  measured
observed  surface  brightness  from  the single  brightest  pixel  and
concluded that  high-redshift (2.2$\la\!z\!\la$3.5) starburst galaxies
have  intrinsic  ultra-violet   (UV)  surface  brightnesses  that  are
typically  4 times  higher than  for low-redshift  starburst galaxies.
Both   M97   and   W98   measured   their   surface   brightness   for
spectroscopically  confirmed   galaxies  in  the   Hubble  Deep  Field
\citep[HDF;][]{will96}.

We  have  measured the  surface  brightness  of  starburst regions  at
$3\!\la\!z\!\la\!6$, using photometry from the Hubble Ultra Deep Field
(HUDF) images \citep{beck06} and  redshifts from the GRism ACS Program
for    Extragalactic   Science    (GRAPES)   project    \citep[PI   S.
Malhotra;][]{pirz04}.  We combine these with earlier published results
comparing  $z\!\simeq\!3$ and  $z\!\simeq\!0$  starbursts (M97,  W98).
The    starburst   intensity   limit    of   starburst    regions   at
$5\!\la\!z\!\la\!6$ sample  is consistent with  that at $z\!\simeq\!3$
and  $z\!\simeq\!0$ to  within the  uncertainties, which  are  about a
factor of three.  \emph {These  high redshift star forming regions are
  thus starbursts,  with a star  formation intensity similar  to their
  local  counterparts  despite any  effects  of differing  metallicity
  and/or  galaxy  size.}  The  starbursts  should  then  be a  set  of
standard surface brightness objects, and can be used to apply Tolman's
test  for  expansion of  the  universe  \citep{tolm30,tolm34} over  an
unprecedented   redshift  range   ($0\!<\!z\!\la\!6$).    Our  surface
brightness  observations  fully  support standard  expanding  universe
models.  This result is robust to even rather large systematic errors,
thanks to the wide redshift range spanned by the data.

This paper is organized as  follows: In \secref{data} we summarize the
HUDF and the GRAPES observations, and we present details of the sample
selection.  In  \secref{m97} we describe our data  analysis to measure
the  UV  spectral  slope  and  the  effective  surface  brightness  of
starburst galaxies  following the method used by  M97, in \secref{w98}
we  apply the  pixel  based method  of  W98 to  estimate the  limiting
surface brightness  for our  galaxies, in \secref{results}  we discuss
measurement   biases   that  will   affect   our  surface   brightness
measurements, and  our results of  the starburst intensity  limit, the
size evolution  \& the change in  the UV spectral  slope.  Finally, in
\secref{summary} we summarize our results.

Throughout this paper we denote the \emph{HST}/ACS F435W, F606W, F775W
and F850LP  filters as \bb,  \vv, \ii, \zz, \emph{HST}/WFPC2  F814W as
\iw\ts and  \emph{HST}/NIC3 F110W and  F160W as \jj\ts  and \hh-bands,
respectively.  We assume a \emph{Wilkinson Microwave Anisotropy Probe}
(WMAP)  cosmology   of  $\Omega_m$=0.24,  $\Omega_{\Lambda}$=0.76  and
\Ho=73~km~s$^{-1}$~Mpc$^{-1}$, in  accord with the recent  3 year WMAP
estimates of \citet{sper07}.  This implies  an age for the Universe of
13.7~Gyr.  Magnitudes are given in the AB system \citep{oke83}.


\section{Observational  Data and Sample Selection}\label{data}

The HUDF is a 400 orbit survey of a $3.4'\times3.4'$ field carried out
with   the  ACS   in  the   \bb,  \vv,   \ii\ts  and   \zz\ts  filters
\citep[see][for  further details]{beck06}.  We  have carried  out deep
unbiased slitless  spectroscopy of  this field with  the ACS  grism as
part of  the GRAPES  project, which was  awarded 40 HST  orbits during
Cycle  12 (ID  9793; PI  S.  Malhotra).   The grism  observations were
taken at four different orientations  in order to minimize the spectra
contamination and overlapping from  nearby sources.  We have extracted
useful low resolution spectra ($R\simeq100$) from 5900\Ang to 9500\Ang
for many sources in the HUDF  down to a limiting magnitude of $z'_{\rm
AB}\!\simeq$27.5 in the AB  system.  Details of the observations, data
reduction  and  final GRAPES  catalog  are  described  in a  paper  by
\citet{pirz04}.

We identify  high-redshift galaxies  on the basis  of their  ACS grism
spectra.  This  identification was based on detecting  the Lyman break
in the  continuum or the  Ly$\alpha$ emission for these  sources. With
the ACS  grism low-resolution  spectra, we are  able to  determine the
redshifts  to an  accuracy  of $\Delta  z\!\approx$0.15  even for  the
faintest   detectable   Lyman    Break   Galaxies   (LBGs)   ($z'_{\rm
  AB}\!\simeq$27.5).  Details  of the selection  process are described
in   \citet{malh05}.    There   are   47  star-forming   galaxies   at
$z\!\simeq\!5\!-\!6$ in  the GRAPES/HUDF with  confirmed spectroscopic
redshifts.  These redshifts are sufficiently high that rest-frame flux
measurements are  available for UV wavelengths comparable  to those at
which M97 and W98 measured the surface brightness of starburst regions
at   $z\!\lesssim\!3$.   We   used  the   following  spectroscopically
confirmed samples for our analysis.
\begin{itemize}

\item   The   $z\!\simeq\!3$  LBGs   from   the   Hubble  Deep   Field
\citep[HDF,][]{giav96,stei96a,stei96b}   were  used  to   compare  our
measurements with  M97.  This sample of 10  galaxies at $z\!\simeq\!3$
forms  the subset  of UV  samples  used by  M97.  We  measured the  UV
spectral  slopes ($\beta$)  using observed  (\vv--\iw)  and ($G$--$R$)
colors,  and  rest-frame UV  fluxes  were  derived  from the  observed
$R$-band   (combined   \vv\ts    and   \iw\ts\ts   light)   magnitudes
\citep{giav96,stei96a,stei96b}.

\item The $z\!\simeq\!4$ \bb-band  dropout galaxies were selected from
the VLT redshift  catalog of \citet{vanz06}.  This sample  is small (4
galaxies),  but  is  useful  for  comparing  $z\!\simeq\!3\!-\!4$  and
$z\!\simeq\!5\!-\!6$ galaxy samples.   We used the observed (\ii--\zz)
color to estimate $\beta$, and  rest-frame UV fluxes were derived from
the  observed \zz-band  magnitudes.   The observed  \ii- and  \zz-band
magnitudes  for $z\!\simeq\!4$  galaxies were  obtained from  the HUDF
catalogs of \citet{beck06}.

\item We  use a sample  of 47 $z\!\simeq\!5\!-\!6$  starburst galaxies
having GRAPES redshifts (Malhotra et  al. 2005; Rhoads et al. 2007, in
preparation) in the HUDF. \tabref{tab1} shows properties (coordinates,
magnitudes,  sizes and redshifts)  for these  47 galaxies.  We further
selected a  subset of 19  galaxies covered by the  \citet{thom05} HUDF
NICMOS  images.   We estimate  $\beta$  from  the observed  (\jj--\hh)
color.  The rest-frame UV fluxes were derived from the observed \jj\ts
and \hh-band magnitudes for 19  galaxies, while the average $\beta$ is
used to  predict \jj\ts and  \hh-band magnitudes for the  remaining 28
galaxies.
\end{itemize}

\section{Starburst Intensity Limit Using M97 Approach}\label{m97}


\subsection{Magnitudes and Color Measurements}\label{mags}

All measurements for galaxies at $z\!\simeq\!4\!-\!6$ were done on the
HST   ACS   HUDF  images   \citep{beck06}   and   HST  NICMOS   images
\citep{thom05}.  The  HST NICMOS images  were reprocessed by  L. Eddie
Bergeron  (private  communication).    The  HST  NICMOS  images  cover
$\sim$50\% of the HUDF ACS field. Therefore, only half of our galaxies
have \jj- and \hh-band imaging.   We need \jj\ts and \hh\ts magnitudes
for galaxies at $z\!\simeq\!5\!-\!6$ to measure the UV spectral slopes
and  luminosities at  $\sim$2200\Ang\ rest-frame.   We  use rest-frame
$\sim$2200\Ang\ts  to  measure the  effective  surface brightness  for
consistency  with  M97.  All  broad-band  magnitudes  are measured  as
\texttt{SExtractor}   \citep{bert96}   \texttt{MAG\_AUTO}  magnitudes,
using dual-image  mode to generate aperture matched  catalogs.  We use
the  \zz-band  image   as  the  detection  image.   \texttt{MAG\_AUTO}
apertures  are Kron-like  \citep{kron80} flexible  apertures  and they
enclose most of  the flux for an object.  These  apertures are same in
all filters for a  given object.  We also measured \texttt{SExtractor}
\citet{petr76}   magnitudes  (\texttt{MAG\_PETRO}),   with  $\eta$=0.2
\citep{holw05}, and isophotal magnitudes (\texttt{MAG\_ISO}).  We find
that   the   average   difference   between   \texttt{MAG\_AUTO}   and
\texttt{MAG\_PETRO}  is $\sim$0.1  mag, while  the  average difference
between  \texttt{MAG\_AUTO} and  \texttt{MAG\_ISO}  is $\sim$0.2  mag.
The  effect of this  magnitude uncertainty  on the  surface brightness
measurements  is  very small  ($<$0.1  dex).   Our  method to  measure
magnitudes is  the same for all galaxies  at $z\!\simeq\!4\!-\!6$, and
is consistent with the  curve-of-growth method used for our comparison
sample of  $z\!\simeq\!3$ galaxies \citep{giav96,stei96a,stei96b}.  We
use \texttt{MAG\_AUTO}  magnitudes to calculate the  UV spectral slope
($\beta$) and the effective surface brightness.

\subsection{The UV Spectral Slope (\boldmath{$\beta$})}\label{beta}

\begin{figure}
\epsscale{0.7}
\plotone{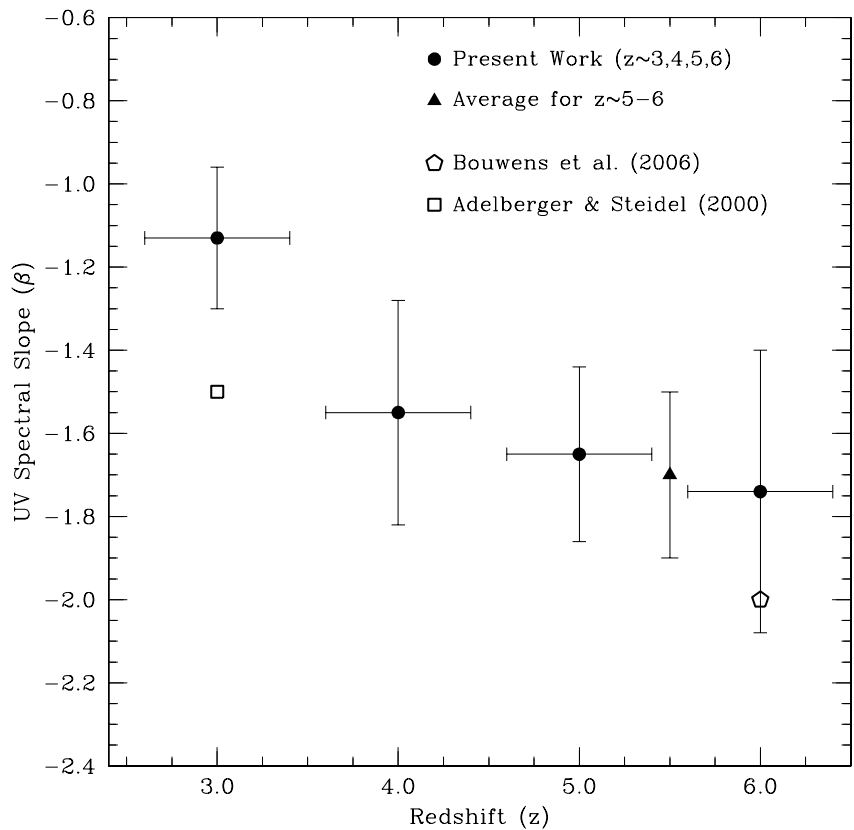}
\caption{UV  spectral  slopes ($\beta$)  v/s  redshift relation.   Mean
  $\beta$  are  plotted  with   error  bars  indicating  the  standard
  deviation of the  mean (i.e.  $\sigma / \sqrt  N$).  We have plotted
  the  data  points  at  $z\!\simeq\!3$  from  \citet{adel00}  and  at
  $z\!\simeq\!6$ from \citet{bouw06} for comparison.}\label{fig1}
\end{figure}

The UV  spectral slope ($\beta$) is  determined from a  power-law fit to
the UV continuum spectrum \citep{calz94},
\[ 
f_{\lambda} \varpropto \lambda^{\beta} ~~,
\]
where  $f_{\lambda}$ is  the flux  density per  unit  wavelength (ergs
s$^{-1}$ cm$^{-2}$ \Ang$^{-1}$).  Converting this into magnitude units
yields   a   linear   relationship   between   $\beta$   and   colors.
\figref{fig1}  shows the  average values  of the  UV  spectral slopes,
$\beta$,     for      galaxies     at     $z\!\simeq\!3\!-\!4$     and
$z\!\simeq\!5\!-\!6$.   The mean  values are  plotted with  error bars
indicating  the  standard deviation  of  the  mean  (i.e.  the  sample
standard deviation ($\sigma$) divided by the square root of the sample
size  ($N$)).  The  UV slopes  for 10  galaxies at  $z\!\simeq\!3$ are
obtained  using the  following two  equations (M97)  for  two slightly
different samples:
\[ 
\beta = 2.55 \cdot (G - R) - 2  \; \; {\rm and} \; \; \beta = 3.23 \cdot (V - I) - 2  ~~.
\]
Here $G$ and $R$ filters  are defined in \citet{stei93}.  We also plot
the average $\beta$ measured for 4 galaxies at $z\!\simeq\!4$, using:
\[
\beta =  5.65 \cdot (i' - z') - 2
\]
where  we   have  used  pivot  wavelengths  for   \ii-  and  \zz-band,
$\lambda_i$=7693\Ang  and  $\lambda_{z'}$=9055\Ang,  respectively,  to
obtain  the  slope  of  5.65 in  $\beta$--color  linear  relationship.
\figref{fig1}  also  shows the  average  $\beta$  for  19 galaxies  at
$z\!\simeq\!5\!-\!6$ obtained using
\[ 
\beta = 2.56 \cdot (J - H) - 2 
\]
where  we   have  used  pivot  wavelengths  for   \jj-  and  \hh-band,
$\lambda_J$=11200\Ang\ts  and $\lambda_H$=16040\Ang,  respectively, to
obtain the slope of 2.56 in $\beta$--color linear relationship.  We do
not  use  (\zz--\jj)  color   to  estimate  $\beta$  for  galaxies  at
$z\!\simeq\!5\!-\!6$ because, the (\zz--\jj) colors can be insensitive
to rest-frame UV colors due to the shorter color baseline and are also
more sensitive to uncertainties in the optical to infrared zero points
\citep{bouw06}.   Therefore for comparison,  we also  measured $\beta$
for galaxies at $z\!\simeq\!5$  using (\ii--\zz) colors and found that
the  average $\beta$ is  --1.53$\pm$0.38 compared  to --1.65$\pm$0.21,
the average  $\beta$ at $z\!\simeq\!5$ using  (\jj--\hh) colors.  This
comparison  implies  that  small   variations  in  the  UV  rest-frame
wavelength and  observed colors does  not affect the slope  within the
quoted uncertainties.

\figref{fig1}   shows  that   the  average   $\beta$   decreases  from
--1.13$\pm$0.17    at    $z\!\simeq\!3$    to    --1.74$\pm$0.35    at
$z\!\simeq\!6$.   The  change in  the  average  slope  shows that  the
galaxies   at   $z\!\simeq\!5\!-\!6$   are   bluer   than   those   at
$z\!\simeq\!3$,   but  redder   than   the  flat   slope  in   $f_\nu$
($f_{\lambda}  \propto \lambda^{-2}$),  that would  be expected  for a
dust-free   starburst   galaxy. 


\subsection{Half-Light Radius (\boldmath{$r_e$}) Measurements}

The half-light radius is defined  as the radius containing 50\% of the
total  flux  of an  object.   The  half-light  radii for  galaxies  at
$z\!\simeq\!3$  are derived by  \citet{giav96} \&  \citet{stei96b} and
are measured such  that half of the total  emission from the starburst
is   enclosed    within   a   circular   aperture.     We   used   the
\texttt{SExtractor} half-light radii (R50, radii enclosing 50\% of the
flux  within a  circular  aperture) obtained  from  the HUDF  \zz-band
catalog  \citep{beck06}  for  galaxies at  $z\!\simeq\!4\!-\!6$.   All
radii are converted from arcsecs to  kpc using a WMAP cosmology in the
cosmological calculator  by \citet{wrig06}.  The  half-light radii for
galaxies  at  $z\!\simeq\!4\!-\!6$   are  measured  at  rest-frame  UV
wavelengths  $\lambda$=1500$\pm$300\Ang.   The  sizes  do  not  change
appreciably when measured within this wavelength range.  \figref{fig2}
shows  the average  values of  the half-light  radii for  10 starburst
galaxies  at $z\!\simeq\!3$, 4  starburst galaxies  at $z\!\simeq\!4$,
and  47 starburst galaxies  at $z\!\simeq\!5\!-\!6$.   Mean half-light
radii are plotted with error bars indicating the standard deviation of
the mean.  The solid and dashed  curves show the trend if sizes evolve
as  $H^{-1}(z)$ or  $H^{-2/3}(z)$, respectively,  where $H(z)$  is the
Hubble parameter  at redshift $z$.   The curves are normalized  to the
mean  size  we  measure  at  $z\!\simeq\!4$  ($\sim$0.21\arcsec\ts  or
$\sim$1.5  kpc).  Comparison  between galaxies  at  $z\!\simeq\!3$ and
$z\!\simeq\!5\!-\!6$  shows that the  galaxy sizes  increase as  we go
from $z\!\simeq\!6$ to $z\!\simeq\!3$.

We independently measured various flavors of \texttt{SExtractor} radii
(half-light,     \citet{petr76},     Kron)     for     galaxies     at
$z\!\simeq\!5\!-\!6$ to assess  the differences in these measurements.
The Petrosian  radius is  defined as the  radius at which  the surface
brightness  is   certain  factor  ($\eta$)  of   the  average  surface
brightness within this  isophote, while the Kron radius  is defined as
the typical  size of the  flexible aperture computed from  the moments
\citep[see     \texttt{SExtractor}      manual     by][for     further
details]{holw05}.   The  difference  in  the  average  values  of  the
half-light    and   the    other   two    radii    was   approximately
$\pm$0.04\arcsec. Therefore,  we expect about 30\%  uncertainty in the
measurements    of   the    half-light   radii    for    galaxies   at
$z\!\simeq\!5\!-\!6$.


\subsection{Calculation of Surface Brightness for Starburst Galaxies}\label{calculations}

We measure the effective  surface brightness for 14 starburst galaxies
at    $z\!\simeq\!3\!-\!4$    and    47    starburst    galaxies    at
$z\!\simeq\!5\!-\!6$ by  adopting the method  used by M97.   First, we
need to  estimate dust  extinction A$_{\rm 1600}$  and $k$-corrections
from the  UV spectral slope  ($\beta$).  Dust extinction  is estimated
using the linear empirical relation between A$_{\rm 1600}$ and $\beta$
\citep{meur99}, which is given by following equation:
\[	
{\rm A}_{\rm 1600}=4.43+1.99 \cdot (\beta) 
\]
where A$_{\rm  1600}$ is the net  absorption in magnitudes  by dust at
1600\Ang.

The next step in correcting apparent flux for corresponding rest-frame
UV flux  is to  apply, where appropriate,  the $k$-correction.  We use
following equation to estimate $k$-correction (M97):
\[ 
k = \frac{f_{\rm 2320}}{f_{\lambda_c/1+z}} = \left [\frac{(1+z) \cdot 2320 {\rm \Ang}}{\lambda_c} \right ]^\beta 
\]
where $\lambda_c$ corresponds to the central wavelength of the filters
used for the observed flux.  Here we reference all observations to the
observations in M97, which has  UV central wavelength of 2320\Ang.  We
apply above mentioned dust  and $k$-corrections to apparent magnitudes
to estimate absolute magnitudes  and intrinsic UV luminosities for our
samples of  $z\!\simeq\!3\!-\!6$ galaxies.  The ratio  of intrinsic UV
to bolometric  luminosity can be calculated for  young starbursts.  By
using the stellar population  models of \citet{bruz03} to convert from
intrinsic F220W flux/luminosity to bolometric luminosity, we find that
the ratio of the UV to bolometric luminosity changes due to variations
in  the metallicity and  the dust  attenuation.  The  luminosity ratio
spans a range  from $\sim$0.2 to $\sim$0.5.  The  adopted ratio of the
intrinsic UV to bolometric luminosity is:
\[
\frac{L}{L_{\rm bol}} \simeq 0.33
\]	
We are using this value for  the UV to bolometric luminosity ratio for
two  reasons: (1)  this bolometric  correction  is very  close to  the
average value we  get from our stellar population  models and (2) this
correction factor is also predicted by the models used by M97.

From  $L_{\rm bol}$  and the  half-light radii  ($r_e$),  we calculate
effective surface brightness (S$_{e}$) using following relation:
\[
{\rm S}_e = \frac{L_{\rm bol}}{2 \pi r_e^2}  \; \; \;  \left (\frac{L_{\sun}}{{\rm kpc}^2} \right ) ~~.
\]	
Here $r_e$ is measured in  kpc and $L_{\rm bol}$ in solar luminosities
($L_{\sun}$).

To  characterize the  S$_{e}$  distribution for  all  galaxies in  our
sample  we consider  the median  and 90$^{th}$  percentiles,  which we
denote as S$_{e,\rm 50}$  and S$_{e,\rm 90}$, respectively.  The upper
limit  to  the  surface  brightness  (starburst  intensity  limit)  of
starbursts is  traced by  S$_{e,\rm 90}$.  Here  we have  used average
$\beta$ (from  \figref{fig1}) for each sample to  estimate the surface
brightness.  \figref{fig3} shows $L_{\rm bol}$ and S$_e$ as a function
of   $r_e$   for  the   $z\!\simeq\!0\!-\!6$   galaxy  samples.    The
$z\!\simeq\!0$  and $z\!\simeq\!0.4$  surface  brightness measurements
are  taken from  M97.  The  $z\!\simeq\!0$  data point  is the  median
measurement for 11 nearby  galaxies.  The S$_{e,\rm 50}$ and S$_{e,\rm
  90}$ surface brightness levels of the combined sample are plotted as
dashed and dotted lines respectively.

The  top  panel  of   the  \figref{fig3}  shows  that  the  bolometric
luminosities for galaxies at $z\!\simeq\!5\!-\!6$ are smaller than the
luminosities  of galaxies  at $z\!\simeq\!3$.   Using a  two-sided K-S
test on these luminosity  distributions, we reject the hypothesis that
the  $z\!\simeq\!3$ and  $z\!\simeq\!5\!-\!6$  luminosities are  drawn
from  the same  population at  $>$99\% probability.   From  the bottom
panel of \figref{fig3},  it is apparent that S$_e$  shows little or no
dependence on $r_e$  over about one order of  magnitude in size; hence
there  is no  dependence on  $L_{\rm bol}$  over about  two  orders of
magnitude  in luminosity.  \figref{fig4}  shows the  effective surface
brightness (S$_{e}$)  as a  function of redshift.   The $z\!\simeq\!0$
and  $z\!\simeq\!0.4$ surface brightness  measurements are  taken from
M97.   From \figref{fig3}  and \figref{fig4},  we find  that S$_{e,\rm
  90}$ of the starbursts  remains constant (within uncertainties) with
redshift.


\section{Starburst Intensity Limit Using W98 Approach}\label{w98}

We  also studied the  surface brightnesses  using the  brightest pixel
approach pioneered by \citet{weed98}. \citet{weed98} measures observed
surface   brightness  of   the   brightest  pixel   for  galaxies   at
$2.2\!\la\!z\!\la\!3.5$  in  the  HDF   and  compared  with  the  local
starbursts by  fading their observed  ($f_\lambda$) surface brightness
by  (1+$z$)$^{-5}$.    We  use   this  approach  to   compare  surface
brightnesses of the brightest pixel for galaxies at $z\simeq\!3\!-\!6$.

A   postage   stamp  (51$\times$51   pixels)   for   each  galaxy   at
$z\simeq5\!-\!6$  was excised  from  the \zz-band  HUDF image.   Using
\zz-band  segmentation maps,  only object  pixels were  selected.  For
each  object pixel  in the  postage  stamp, we  estimate the  apparent
magnitude.  For  the brightest  pixel in each  galaxy, the  average UV
spectral slope $\beta$  was used to predict \jj-band  (for galaxies at
$z\!\simeq\!5$)   and  \hh-band   (for  galaxies   at  $z\!\simeq\!6$)
magnitudes from \zz-band  apparent magnitudes.  Using similar approach
as discussed in \secref{calculations},  we calculated the intrinsic UV
luminosity  ($L_{\sun}^{\rm  UV}$) for  the  brightest  pixel in  each
galaxy at $z\!\simeq\!5\!-\!6$.   For surface brightness measurements,
we divide  the intrinsic  UV luminosity  by the area  of one  pixel in
kpc$^2$.

\figref{fig5}     shows     surface     brightness     ($L_{\sun}^{\rm
UV}$~kpc$^{-2}$) for the brightest pixel in each of the 47 galaxies at
$z\!\simeq\!5\!-\!6$ and  4 galaxies at $z\!\simeq\!4$.   We have also
plotted 18 galaxies at $z\!\simeq\!3$  from W98.  The W98 galaxies has
observed surface brightnesses for  the brightest pixels and hence, for
proper comparison, we converted  observed surface brightnesses for W98
galaxies to their corresponding rest-frame surface brightnesses.  Here
we have used  average $\beta$ for $z\!\simeq\!3$ galaxies  as shown in
\figref{fig1}, to estimate  extinction and $k$-correction.  The median
(S$_{bp,\rm 50}$)  and 90$^{th}$ percentile  (S$_{bp,\rm 90}$) surface
brightness levels  of the  combined sample are  plotted as  dashed and
dotted lines,  respectively.  Here,  the starburst intensity  limit of
starbursts  is traced  by  S$_{bp,\rm 90}$.   The  S$_{bp,\rm 90}$  of
galaxies at $z\!\simeq\!3$ and  $z\!\simeq\!5\!-\!6$ is same to within
a factor of $\sim$0.10 dex.

We cannot  properly compare  the brightest pixel  surface brightnesses
between   the   W98  sample   at   $z\!\simeq\!0$   and  galaxies   at
$z\!\simeq\!3\!-\!6$ for  two reasons.  First, the aperture  sizes for
W98  galaxies at  $z\!\simeq\!0$ are  larger than  the  physical sizes
corresponding to one pixel  at $z\!\simeq\!3\!-\!6$, and the sizes for
W98  galaxies at  $z\!\simeq\!0$ are  also larger  than  the effective
radii  measured by \citet{meur95}  for some  of these  local galaxies.
Second,  the discrepancy  between the  W98 adopted  UV  spectral slope
($\beta$) and $\beta$  from \citet{meur95} is large for  some of these
galaxies, which  affects the  applied extinction for  their luminosity
measurements.

The effective  surface brightness  approach of M97  (\secref{m97}) and
the  brightest  pixel approach  of  W98  suggests  that the  starburst
intensity limit, as  defined by S$_{e,\rm 90}$ or  S$_{bp,\rm 90}$, of
the    starbursts   is    unchanged   (within    uncertainties)   from
$z\!\simeq\!5\!-\!6$ down to $z\!\simeq\!3$.


\section{Results and Discussion}\label{results}

\subsection{Selection and Measurement Effects}\label{bias}

\emph{Different radii-measurements.}  --- The results shown in Figures
2  and   4  use   \texttt{SExtractor}  half-light  radii   derived  by
\citet{beck06}.   Here  we  discuss  the uncertainty  in  the  surface
brightness  measurements  due to  radii  measurements.   We use  three
different  flavors of  radii measured  using  \texttt{SExtractor}.  We
measured the  half-light radii, the \citet{petr76} radii  and the Kron
radii for galaxies at $z\!\simeq\!5\!-\!6$.  We found that the average
difference  between the  half-light  radius and  other  two radii  was
approximately $\pm$0.04\arcsec.   This difference does  not affect our
conclusion that $z\!\simeq\!5\!-\!6$  galaxies are smaller compared to
$z\!\simeq\!3$ galaxies.  As shown in \figref{fig2}, the average value
of  the  half-light  radii  for galaxies  at  $z\!\simeq\!5\!-\!6$  is
0.14\arcsec\ts compared  to 0.32\arcsec\ts which is  the average value
for galaxies  at $z\!\simeq\!3$.   The uncertainty in  three different
radii  measurements  is   $\pm$0.04\arcsec.   Therefore  for  a  given
luminosity, we  expect that this difference in  radii measurement will
cause $\sim$0.2 dex difference in S$_{e,\rm 90}$ estimate.
 
\emph{Surface  brightness   selection.}   ---  The   limiting  surface
brightness  in  the  HUDF  samples  is S$_{e}  \sim$  5.0  x  10$^{9}$
$L_{\sun}$ kpc$^{-2}$  at $z\!\simeq\!5\!-\!6$, which  is much fainter
than the observed maximum surface brightness (S$_{e,\rm 90} \simeq 2.3
{\rm \;  x \;} 10^{11}$  $L_{\sun}$ kpc$^{-2}$) for all  galaxies.  At
low surface brightness limit our sample could be incomplete but at the
maximum surface brightness level our sample is complete.

\emph{Size  selection effect.}   --- The  $z\!\simeq\!3$  galaxies are
spectroscopically  confirmed  by   Keck  observations.   The  limiting
magnitude for Keck spectroscopy  is $R \lesssim 25.3$ \citep{stei96b}.
Using  S$_{e,\rm 90}$  and  the limiting  magnitude,  we estimate  the
minimum  observable   size  for  this   sample  as  $\sim$700   pc  or
$\sim$0.1\arcsec.    Therefore,    this   sample   of    galaxies   at
$z\!\simeq\!3$ is not biased against smaller sizes.

\emph{Flux uncertainty.}   --- Here we discuss  three possible sources
of uncertainties in the magnitudes (for a given size) that will affect
our  surface brightness  measurements. They  are as  follows:  (1) The
average \texttt{SExtractor}  uncertainties in $J$,  $H$ magnitudes are
$\sim$0.2  mag.  The  largest  magnitude uncertainties  in the  sample
range up  to $\sim$0.5  mag for three  objects.  Even this  worst case
magnitude uncertainty  affects S$_{e,\rm  90}$ by only  $\sim$0.2 dex.
The  uncertainty  in S$_{e,\rm  90}$  due  to  the average  difference
between  various  magnitudes (\texttt{MAG\_AUTO},  \texttt{MAG\_PETRO}
and  \texttt{MAG\_ISO})  is very  small  ($<$0.1  dex)  and hence  the
magnitude  uncertainty dominates  the uncertainty  in  S$_{e,\rm 90}$.
(2) The  ratio  of  the  UV  to bolometric  luminosity  for  starburst
galaxies is based on the stellar population models.  We find that this
ratio is affected  by the change in metallicity  or dust extinction in
the stellar population models. The uncertainty in this ratio can be as
large as  a factor of  $\sim$2.0.  This uncertainty in  the bolometric
correction will  change S$_{e,\rm 90}$  estimate by $\sim$0.3  dex but
this is systematic uncertainty and will affect all starburst galaxies.
(3) We  have used  linear fit  to  the observed  relation between  the
F$_{\rm FIR}$/F$_{\rm  1600}$ and the  UV spectral slope  ($\beta$) to
estimate  A$_{\rm 1600}$ \citep[Fig.1  in][]{meur99}.  The  scatter in
this plot can cause an uncertainty in A$_{\rm 1600}$ of $\sim$0.4 mag.
This uncertainty in A$_{\rm 1600}$ will change S$_{e,\rm 90}$ estimate
by  $\sim$0.2 dex.   This uncertainty  will systematically  affect all
galaxies in this study.

Therefore, if  these logarithmic  uncertainties add in  quadrature, we
would expect the  total uncertainty in S$_{e,\rm 90}$  to be $\sim$0.5
dex.

\subsection{The Starburst Intensity Limit}

We measure the effective  surface brightness (S$_e$) and the brightest
pixel  surface brightness  (S$_{bp}$) of  the \emph{spectroscopically}
confirmed  galaxies at  $z\!\simeq\!5\!-\!6$ in  the HUDF  and compare
with the spectroscopically confirmed galaxies at $z\!\simeq\!3$ in the
HDF.  We  conclude that to  better than a  factor of 3,  the starburst
intensity    limit    of     starbursts    at    $z\!\simeq\!3$    and
$z\!\simeq\!5\!-\!6$  are  the same.   Using  the  K-S  test on  these
distributions, the resulting probabilities ($\gtrsim$20\%) support the
hypothesis  that   these  distributions   are  drawn  from   the  same
population.   By  combining the  samples  at $z\!\simeq\!3\!-\!4$  and
$z\!\simeq\!5\!-\!6$, we find a mean  S$_{e,\rm 90} \simeq 2.3 {\rm \;
  x  \;}  10^{11}$  $L_{\sun}$  kpc$^{-2}$  (with a  factor  of  3  or
$\sim$0.5 dex uncertainty).  We quantify  the scatter in the S$_e$ and
S$_{bp}$  distributions  (Figure  5   and  6)  by  measuring  standard
deviation ($\sigma$)  from a Gaussian fit to  these distributions.  We
find    that     the    scatter    in     the    S$_e$    distribution
($\sigma_{log(S_e)}\!\simeq\!0.37$) is higher  than the scatter in the
S$_{bp}$                                                   distribution
($\sigma_{log(S_{bp})}\!\simeq\!0.27$). Therefore, the brightest pixel
method  could be  very useful  in estimating  the  starburst intensity
limit of starburst galaxies.

The  approximate   constancy  of  peak  starburst   intensity  can  be
interpreted  as the evidence  that the  interstellar medium  (ISM) can
only  support  some  maximum  pressure.  \citet{heck90}  used  [S  II]
emission  line ratios  to determine  the  ISM pressure  (P$_0$) for  a
sample of  galaxies (mostly  starbursts) undergoing a  strong galactic
wind.   The star  formation  intensity required  to  produce P$_0$  is
S$_{e} \simeq 1.7 {\rm \;  x \;} 10^{11}$ $L_{\sun}$ kpc$^{-2}$ (M97).
This value  agrees closely with  our S$_{e,\rm 90}$ but  derived using
different method.   The physical process can be  explained by assuming
that the  ISM pressure provides  a ``thermostat'' for  star formation,
such  that strong  outflows will  result whenever  the  pressure rises
above P$_0$  and shut  down further star  formation for a  time.  This
results  in  a characteristic  peak  starburst  intensity.  We  convert
S$_{e,\rm 90}$ to an equivalent star formation intensity ($M_{\sun} \;
{\rm yr}^{-1} \; {\rm  kpc}^{-2}$) by using conversion factors between
UV luminosity and star formation  rate from \citet{kenn98} and M97. We
get star formation intensity in  the range of 30--50 $M_{\sun} \; {\rm
yr}^{-1} \; {\rm kpc}^{-2}$ depending on the conversion factor.
This peak intensity is  physically distinct from the minimum intensity
required  to produce  a galactic  wind, which  is orders  of magnitude
lower, at  $\sim 0.1 \; M_{\sun}  \; {\rm yr}^{-1}  \; {\rm kpc}^{-2}$
\citep{lehn96,heck01}.   Thus, {\em  every}  galaxy in  our sample  is
expected to have a wind, as \citet{lehn07} remark for a similar sample
of $z\ga 3$ Lyman break  galaxies.  However, only in galaxies near the
starburst intensity limit does the ``thermostat'' become active.

While the  peak starburst intensity stays constant  with redshift, the
dust  optical depth  that we  infer from  the observed  spectral slope
$\beta$  decreases systematically with  redshift.  This  suggests that
the star formation intensity limit does not depend on dust content, at
least over the range of dust optical depths $\tau_{dust} \ga 1$ probed
by  our  sample.   At  yet  higher  redshifts  we  might  expect  that
$\tau_{dust} <  1$, and  that the starburst  radiation field  would no
longer couple  efficiently to  the interstellar medium.   Whether this
would substantially change the  starburst intensity limit would depend
on the  fraction of  starburst wind driving  that is due  to radiation
pressure  rather  than  stellar   winds  and  mechanical  energy  from
supernova explosions.

The  constancy  of   maximum  star-formation  surface  intensity  with
redshift provides a basis for a strong test of the expanding Universe.
Standard  cosmologies predict that  the bolometric  surface brightness
should  vary  with  redshift  as  $(1+z)^{-4}$.   Our  results  (using
standard cosmology)  combined with M97  results at low  redshifts show
that  the maximum  star formation  intensity remains  constant (within
uncertainties) from $z\!\simeq\!0$ to $z\!\simeq\!6$.  This conclusion
depends  critically on  the use  of a  standard cosmology  to  go from
observed flux and radius  to inferred star formation intensity.  Thus,
if the  peak star formation rate  per unit area is  controlled by some
physical limit  that is based on  local, redshift-independent physics,
our  observations  essentially  require  standard  surface  brightness
dimming.  While other evolutionary  effects could become important, we
have minimized these by measuring surface brightnesses at nearly fixed
rest wavelength.  Had we instead  taken a ``tired light'' model, where
bolometric   surface  brightness   falls  of   as   $(1+z)^{-1}$,  our
observations would require the true star formation intensity to be
dramatically lower at  high redshift--- by  a factor of  order $(1+z)^3
\sim  7^3  \sim  300$.   This  factor greatly  exceeds  the  estimated
uncertainties in our analysis.  Hence,  the wide redshift range of our
sample yields  a strong application  of Tolman's \citep{tolm30,tolm34}
test, and we derive strong evidence in favor of the expanding Universe
and against any alternative ``tired light'' models.

\subsection{Size and Luminosity Evolution}

\begin{figure}
\epsscale{0.7}
\plotone{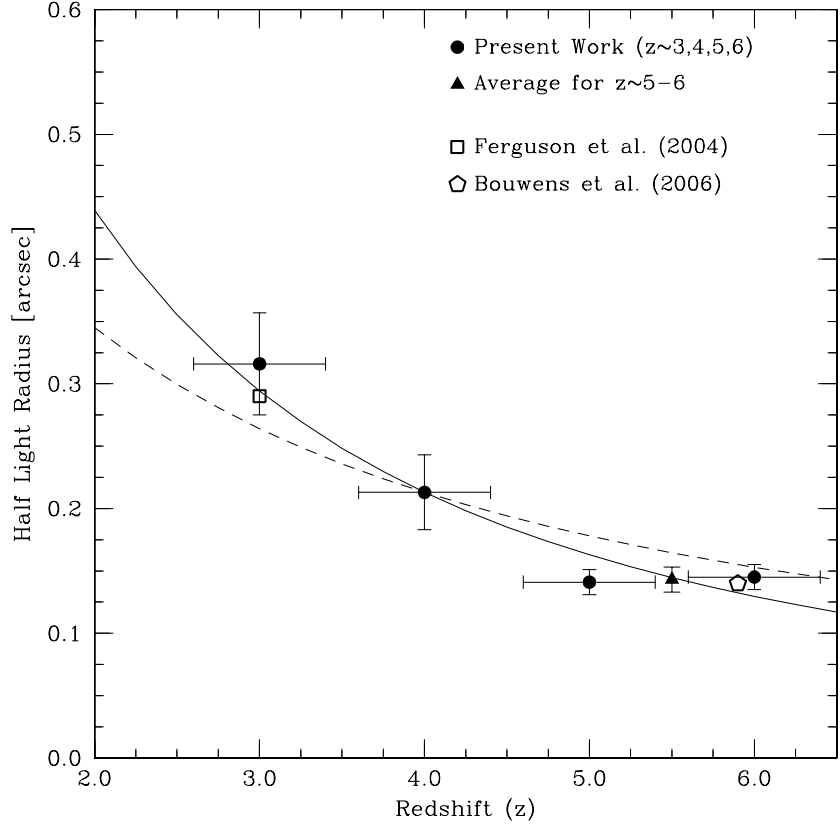}
\caption{Size v/s redshift relation. Mean half-light radii are plotted
  with error bars indicating the  standard deviation of the mean (i.e.
  $\sigma / \sqrt N$).   The solid and dashed curves  shows the trend
  if  sizes  evolve as  $H^{-1}(z)$  and $H^{-2/3}(z)$,  respectively.
  Both  curves  are normalized  to  the  mean  size at  $z\!\simeq\!4$
  ($\sim$0.21\arcsec\ts or  $\sim$1.5 kpc).  We have  plotted the data
  points  at $z\!\simeq\!3$ from  \citet{ferg04} and  at $z\!\simeq\!6$
  from \citet{bouw06} for comparison.}\label{fig2}
\end{figure}

\begin{figure}
\epsscale{0.7}
\plotone{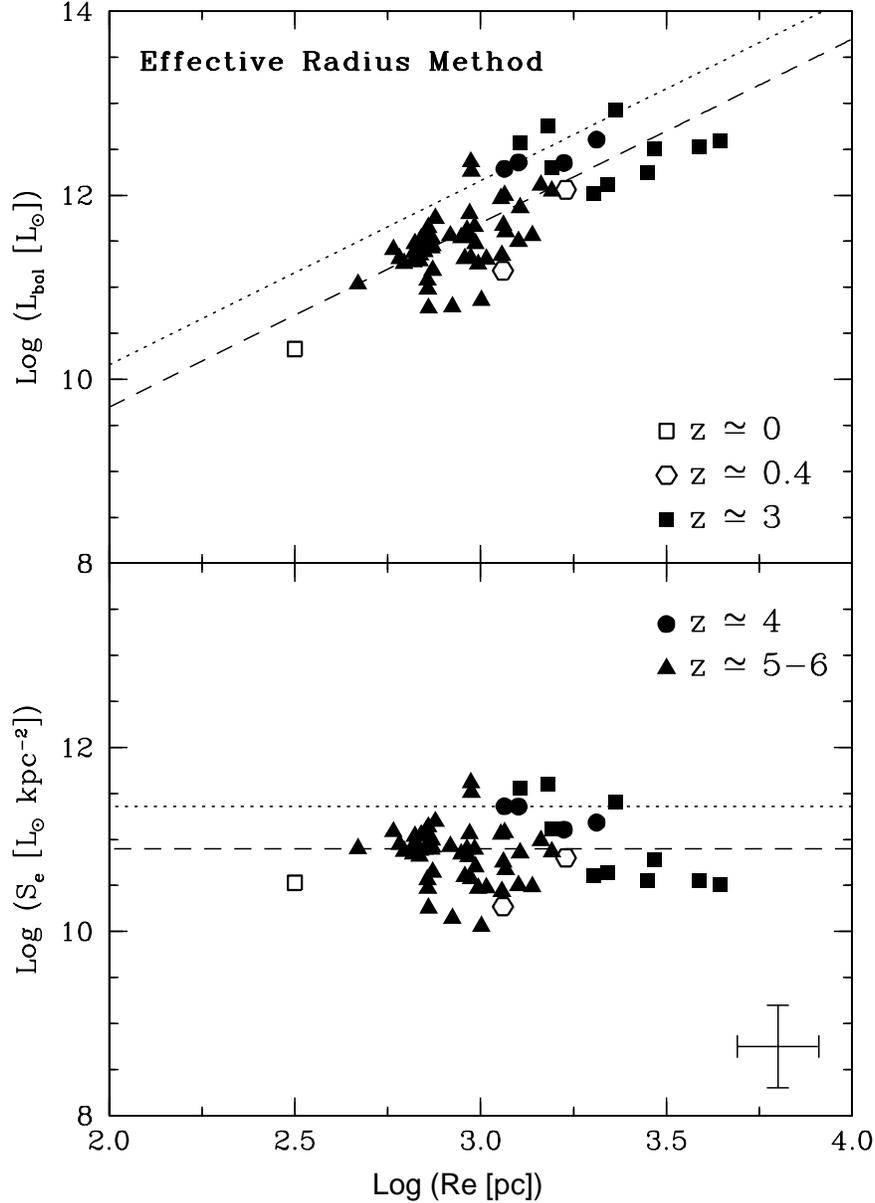} 
\caption{Bolometric  luminosity ($L_{\sun}^{\rm  bol}$)  and effective
  surface   brightness   ($L_{\sun}^{\rm   bol}$~kpc$^{-2}$)   against
  effective radii for starburst galaxies.  The filled squares, circles
  and  triangles  are  measurements  for galaxies  at  $z\!\simeq\!3$,
  $z\!\simeq\!4$  and  $z\!\simeq\!5\!-\!6$,  respectively.  The  open
  square  ($z\!\simeq\!0$)  is the  median  measurement  of 11  nearby
  galaxies from M97.  The open diamonds ($z\!\simeq\!0.4$) are S$_{e}$
  measurements from  M97.  The dotted  and dashed lines  correspond to
  S$_{e,\rm   90}$  and   S$_{e,\rm  50}$   of  the   combine  sample.
  Uncertainties in  $z\!\simeq\!3\!-\!6$ surface brightness  and radii
  measurements are shown  in lower right corner, details  are given in
  \secref{bias}.}\label{fig3}
\end{figure}

\begin{figure}
\epsscale{0.7}
\plotone{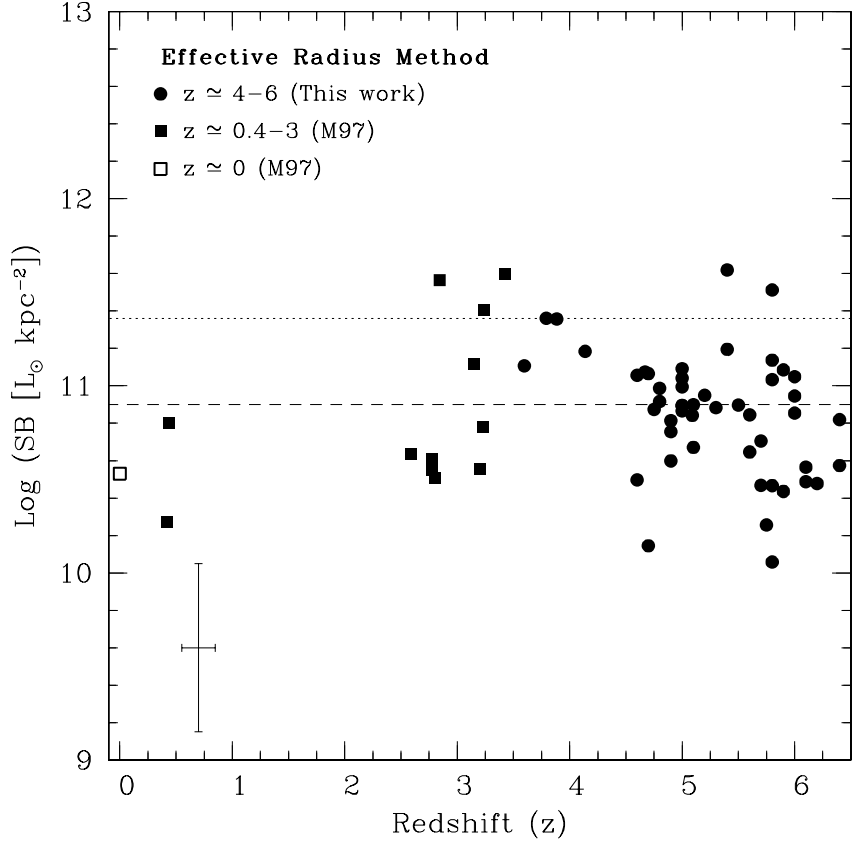}
\caption{Bolometric   effective  surface   brightness  ($L_{\sun}^{\rm
    bol}$~kpc$^{-2}$)  as a  function  of redshift.   The open  square
    ($z\!\simeq\!0$) is  the median measurement of  11 nearby galaxies
    from  M97.   The  filled  squares ($z\!\simeq\!0.4$)  are  S$_{e}$
    measurements  from M97.  The  filled squares  ($z\!\simeq\!3$) are
    the galaxies from the sample  of M97 for which we measured surface
    brightnesses.   The  circles  are   the  galaxies  in  our  sample
    ($z\!\simeq\!4\!-\!6$).  The dotted and dashed lines correspond to
    S$_{e,\rm  90}$   and  S$_{e,\rm  50}$  of   the  combine  sample.
    Uncertainties in  $z\!\simeq\!3\!-\!6$ surface brightness  (due to
    radii  and flux  uncertainties)  is shown  in  lower left  corner,
    details are given in \secref{bias}.}\label{fig4}
\end{figure}

\begin{figure}
\epsscale{0.7}
\plotone{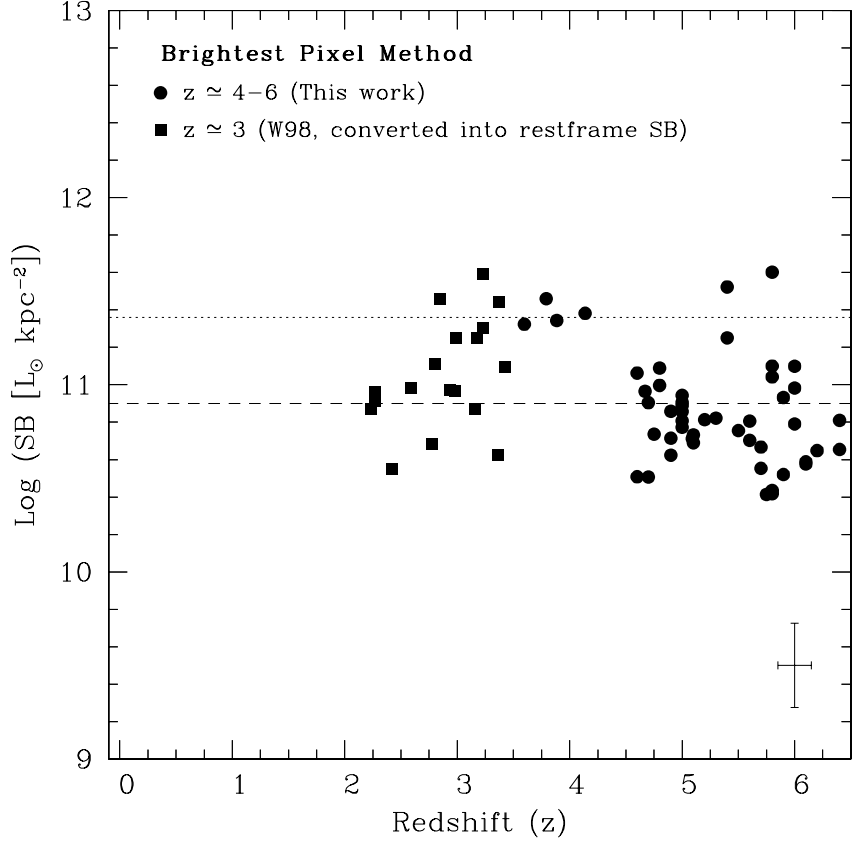}
\caption{Surface Brightness  ($L_{\sun}^{\rm UV}$~kpc$^{-2}$) obtained
  from the brightest pixel of 47 galaxy images at $z\!\simeq\!5\!-\!6$
  and  4 galaxy images  at $z\!\simeq\!4$  .  The  squares are  18 W98
  galaxies.  The  W98 galaxies  had observed surface  brightnesses and
  hence,  for   proper  comparison,  we   converted  observed  surface
  brightnesses to corresponding  rest-frame surface brightnesses.  The
  dotted and dashed lines correspond to S$_{bp,\rm 90}$ and S$_{bp,\rm
    50}$ of  the combine sample.  Uncertainties  in surface brightness
  (due to  flux uncertainties  only) is shown  in lower  right corner,
  details are given in \secref{bias}.}\label{fig5}
\end{figure}

An average galaxy at $z\!\simeq\!5\!-\!6$ in our sample has half-light
radii   of   $\sim$0.8  kpc   or   $\sim$0.14\arcsec,   as  shown   in
\figref{fig2},  which is  in good  agreement with  a number  of recent
studies  \citep{bouw04,bouw06,pirz07,dow07}.   \citet{ferg04} compares
sizes  of galaxies  at $z\!\simeq\!1$--$5$  within a  fixed luminosity
range.  Our results agree with \cite{ferg04} for $z\!\simeq\!3\!-\!4$.
The UV intrinsic luminosities for  our sample of galaxies are brighter
or equal to $L^{*}$ (i.e. $L\!\gtrsim\!L^{*}$) at respective redshifts
but  we  do  not  require  any particular  minimum  luminosity,  while
\citet{ferg04} do  require a  minimum luminosity.  Given  that surface
brightness is near-constant,  a minimum luminosity immediately implies
some  minimum  radius  for  inclusion in  the  \citet{ferg04}  sample.
\figref{fig2} also shows comparison  of our size measurements with the
\citet{ferg04} and  the \citet{bouw06}  results. The solid  and dashed
curves  in  the  \figref{fig2}  show  the trend  if  sizes  evolve  as
$H^{-1}(z)$ or $H^{-2/3}(z)$, respectively, where $H(z)$ is the Hubble
parameter at redshift $z$.  The curves are normalized to the mean size
we measure at  $z\!\simeq\!4$ ($\sim$0.21\arcsec\ts or $\sim$1.5 kpc).
The   galaxy  sizes  from   $z\!\simeq\!3$  to   $z\!\simeq\!6$  scale
approximately  as  the  Hubble  parameter $H^{-1}(z)$,  though  it  is
difficult  to  conclude  with  certainty  how  sizes  scale  from  the
measurements of high redshift ($z\!\gtrsim\!3$) galaxies only, because
two trends diverge significantly at lower redshifts ($z\!<\!3$).

We  also find  that  the  bolometric or  intrinsic  UV luminosity  for
galaxies  at  $z\!\simeq\!5\!-\!6$  is  lower  than  for  galaxies  at
$z\!\simeq\!3\!-\!4$.    The  galaxies  at   $z\!\simeq\!5\!-\!6$  are
smaller  compared  to   $z\!\simeq\!3$  galaxies,  while  the  surface
brightness as measured  in \secref{calculations} remains approximately
constant.   Therefore, the luminosity  evolution could  be due  to the
constant  upper limit  on  the  surface brightness  as  a function  of
redshift.

\subsection{Evolution in UV Spectral Slope \boldmath{($\beta$)}}\label{betainfo}

Our $z\!\simeq\!3\!-\!6$ sample also allows us to place constraints on
the rest-frame  UV slope.  We obtained these  constraints by measuring
rest-frame UV colors  of $z\!\simeq\!3\!-\!4$ and $z\!\simeq\!5\!-\!6$
galaxies  as shown  in \figref{fig1}.   A comparison  of  our measured
colors    with    those    obtained    in   two    previous    studies
\citep{stan05,bouw06}    show   agreement    within    our   1$\sigma$
uncertainties.   The  mean  $\beta$   inferred  from  this  study  for
$z\!\simeq\!6$  galaxies is  $\beta$=--1.74$\pm$0.35, which  is redder
than the  $\beta$=--2.0$\pm$0.3 inferred in the  \citet{bouw06} or the
$\beta$=--2.2$\pm$0.2  inferred in the  \citet{stan05}.  We  also find
that the  mean $\beta$ value  for galaxies at  $z\!\simeq\!5\!-\!6$ is
bluer than  the $\beta$=--1.1$\pm$0.2 we measure  at $z\!\simeq\!3$ or
the    $\beta$=--1.5$\pm$0.4   observed    by    the   \citet{adel00}.
Irrespective  of  the exact  $\beta$,  the  mean  rest-frame UV  slope
observed  at  $z\!\simeq\!5\!-\!6$  is  bluer than  that  observed  at
$z\!\simeq\!3$.  This evolution  is consistent  with number  of recent
studies \citep{stan05, yan05,bouw06}.

To   understand   the  evolution   in   the   $\beta$,   we  use   the
\texttt{STARBURST99}   stellar  synthesis   code   \emph{version  5.1}
\citep{leit99,vazq05}  to  investigate the  variations  in $\beta$  as
function  of IMF,  metallicity  and the  star  formation history.   We
assume  two  different  metallicities  ($Z$  = 0.004  and  0.02),  two
different  star formation histories  (constant and  instantaneous) and
two different versions of  the \citet{salp55} IMFs ($\alpha$=2.35 with
M$_{up}$=100   M$_{\odot}$    and   $\alpha$=2.35   with   M$_{up}$=30
M$_{\odot}$).   We measure  $\beta$ for  various models  by  fitting a
first-order  polynomial  to  the  UV spectra  through  the  wavelength
interval 1250--1850\Ang.  We find  that the changes in metallicity and
the IMF have much smaller effect  on the rest-frame UV slope for young
($\lesssim$20   Myr)   starbursts.   Our   results   agree  with   the
\citet{leit99}  models  showing  that   the  UV  spectral  slopes  are
independent of  evolution and IMF effects.   \citet{leit99} shows that
for young  starbursts, the UV spectral slope  ($\beta$) at 2500\Ang\ts
also does not change very much  as a function of model parameters ($Z$
and  IMF).    Therefore  the   observed  evolution  in   $\beta$  from
$z\!\simeq\!5\!-\!6$  to  $z\!\simeq\!3$ is  more  likely  due to  the
change in the dust content.

\section{Summary}\label{summary}

We    have    measured    the    starburst   intensity    limit    for
\emph{spectroscopically  confirmed}  galaxies at  $z\!\simeq\!5\!-\!6$
from  ACS grism  survey GRAPES  in the  HUDF.  We  find that  there is
little  variation in  the  surface brightness  from $z\!\simeq\!3$  to
$z\!\simeq\!6$ and the starburst intensity limit is within a factor of
3 when compared with the sample of $z\!\simeq\!3\!-\!4$ galaxies.  The
constancy  of  starburst intensity  limit  for  starburst galaxies  at
$z\!\simeq\!5\!-\!6$  combined with  the results  obtained by  M97 for
starbursts  at $z\!\lesssim\!3$  implies that  the  physical processes
limiting starburst  intensity at lower  redshifts also apply  to these
high  redshift galaxies.  We  find that  the high  redshift starbursts
have   a  smaller   characteristic  linear   size  than   their  local
counterparts,  and  a correspondingly  lower  luminosity (since  their
surface  brightnesses  are  similar,  and their  sizes  smaller).   We
observe   the   galaxy   size   evolution   from   $z\!\simeq\!3$   to
$z\!\simeq\!6$  and find  that the  sizes scale  approximately  as the
Hubble parameter $H^{-1}(z)$.  Finally,  using rest-frame UV colors we
conclude   that  the  evolution   in  the   UV  spectral   slope  from
$z\!\simeq\!3$ to  $z\!\simeq\!6$ reinforces the  dust evolution which
leads to  bluer galaxies at $z\!\simeq\!5\!-\!6$  compared to galaxies
at  $z\!\simeq\!3$. This  implies that  starbursts were  less obscured
when the universe was younger  and had lower heavy element abundances.
Any future  search for galaxies at  higher redshifts \citep[e.g.\emph{
  James Webb  Space Telescope  (JWST)};][]{wind07} needs to  take into
account  the  size  evolution  \&  constancy  of  surface  brightness;
therefore the decrease in characteristic luminosity with redshift.


\acknowledgments 
We would  like to thank Rogier  Windhorst, Rolf Jansen  and Seth Cohen
for providing  useful comments on  earlier drafts of this  paper. This
work  was supported  by grants  GO 9793  and GO  10530 from  the Space
Telescope  Science Institute,  which is  operated by  AURA  under NASA
contract NAS5-26555.  We also  thank the anonymous referee for helpful
suggestions that improved the paper.


\clearpage

\begin{deluxetable}{ccccccccccc}


\tabletypesize{\scriptsize} 

\tablewidth{0pt}

\tablecaption{Sample of $z\!\simeq\!4\!-\!6$ Galaxies.\label{tab1}}

\tablehead{\colhead{HUDF} & \colhead{RA} & \colhead{DEC} & \colhead{$B$} & \colhead{$V$} & \colhead{$i'$} & \colhead{$z'$} & \colhead{$J$} & \colhead{$H$} & \colhead{R50\tablenotemark{*}} & \colhead{z} \\ 
\colhead{ID\tablenotemark{\dagger}} & \colhead{(J2000)} & \colhead{(J2000)} & \colhead{(mag)} & \colhead{(mag)} & \colhead{(mag)} & \colhead{(mag)} & \colhead{(mag)} & \colhead{(mag)} & \colhead{(pix)} & \colhead{(GRAPES)} } 

\startdata
   119  &  53.1660072  &  --27.8238735  &   28.98  &   29.22  &   27.46  &   27.83  &  \nodata  &  \nodata  &   3.42  &   5.09  \\
   322  &  53.1716057  &  --27.8207884  &   99.00  &   31.94  &   29.15  &   27.05  &  \nodata  &  \nodata  &   5.33  &   5.70  \\
   457  &  53.1627077  &  --27.8189684  &   99.00  &   33.04  &   29.77  &   28.36  &  \nodata  &  \nodata  &   4.00  &   5.80  \\
   865 $\S$ & 53.1652728 & --27.8140614 & 27.48 & 25.47 & 24.71 & 24.51 & \nodata & \nodata & 5.83 & 3.89 $\ddagger$ \\
  1115  &  53.1722701  &  --27.8119757  &   99.00  &   27.93  &   26.38  &   26.19  &   26.71  &   26.21  &   4.67  &   4.70  \\
  1392  &  53.1563542  &  --27.8095882  &   99.00  &   31.18  &   27.78  &   28.44  &  \nodata  &  \nodata  &   2.43  &   5.10  \\
  2225  &  53.1667243  &  --27.8041607  &   99.00  &   29.68  &   26.73  &   25.16  &   25.47  &   25.29  &   5.24  &   5.80  \\
  2285  &  53.1683346  &  --27.8041253  &   99.00  &   29.75  &   27.92  &   27.83  &  \nodata  &  \nodata  &   3.18  &   5.20  \\
  2408  &  53.1885344  &  --27.8034642  &   31.30  &   28.69  &   26.87  &   26.68  &  \nodata  &  \nodata  &   5.87  &   4.90  \\
  2599  &  53.1626591  &  --27.8022980  &   99.00  &   28.83  &   27.20  &   27.12  &   27.56  &   27.80  &   3.81  &   5.00  \\
  2631  &  53.1774780  &  --27.8024502  &   99.00  &   99.00  &   29.82  &   27.88  &  \nodata  &  \nodata  &   4.21  &   6.60  \\
  2690  &  53.1407464  &  --27.8021066  &   99.00  &   33.28  &   29.24  &   27.35  &  \nodata  &  \nodata  &   3.26  &   5.90  \\
  2881  &  53.1415939  &  --27.8005701  &   99.00  &   27.65  &   25.93  &   25.70  &   25.66  &   25.50  &   5.62  &   4.60  \\
  2894  &  53.1462479  &  --27.8008152  &   99.00  &   29.90  &   27.79  &   27.80  &  \nodata  &  \nodata  &   3.64  &   5.30  \\
  2898  &  53.1798235  &  --27.8008762  &   36.61  &   28.51  &   27.00  &   27.15  &   27.66  &   27.75  &   3.74  &   4.80  \\
  3250  &  53.1326635  &  --27.7989462  &   99.00  &   30.39  &   27.45  &   27.59  &  \nodata  &  \nodata  &   4.62  &   4.90  \\
  3317  &  53.1439706  &  --27.7986558  &   31.40  &   29.89  &   28.32  &   27.11  &   27.28  &   27.26  &   7.86  &   6.10  \\
  3325  &  53.1439441  &  --27.7988859  &   99.00  &   31.96  &   28.77  &   27.13  &   27.26  &   27.02  &   3.92  &   6.00  \\
  3377  &  53.1359748  &  --27.7984127  &   32.41  &   35.99  &   28.59  &   27.70  &   26.97  &   26.02  &   4.05  &   5.60  \\
  3398  &  53.1358677  &  --27.7983281  &   31.61  &   31.30  &   28.23  &   26.82  &   27.03  &   27.27  &   4.83  &   5.60  \\
  3450  &  53.1428483  &  --27.7978540  &   32.37  &   33.01  &   28.86  &   27.51  &   27.02  &   27.44  &   6.39  &   5.90  \\
  3503  &  53.1429396  &  --27.7982143  &   99.00  &   31.16  &   29.37  &   27.92  &  \nodata  &  \nodata  &   5.52  &   6.40  \\
  3807  &  53.1457274  &  --27.7966782  &   99.00  &   31.90  &   29.28  &   28.33  &  \nodata  &  \nodata  &   4.10  &   6.10  \\
  3968  &  53.1833352  &  --27.7959542  &   32.56  &   30.25  &   27.83  &   28.72  &  \nodata  &  \nodata  &   4.20  &   4.70  \\
  4050  &  53.1392840  &  --27.7957997  &   31.35  &   31.44  &   29.58  &   27.49  &   27.33  &   27.46  &   3.74  &   6.00  \\
  4173  &  53.1721175  &  --27.7950895  &   99.00  &   29.56  &   27.34  &   27.16  &  \nodata  &  \nodata  &   3.55  &   5.00  \\
  5307  &  53.1908539  &  --27.7903658  &   99.00  &   29.79  &   27.41  &   26.96  &   26.67  &   26.73  &   3.67  &   5.00  \\
  5788  &  53.1456512  &  --27.7882204  &   31.06  &   29.89  &   27.61  &   27.02  &  \nodata  &  \nodata  &   6.07  &   5.10  \\
  6329  &  53.1466545  &  --27.7861315  &   99.00  &   31.16  &   28.10  &   27.02  &   27.15  &   27.46  &   3.97  &   5.50  \\
  6515  &  53.1273683  &  --27.7851701  &   99.00  &   28.55  &   27.34  &   27.59  &  \nodata  &  \nodata  &   3.13  &   4.75  \\
  7050  &  53.1510303  &  --27.7828664  &   31.22  &   29.48  &   27.45  &   26.89  &   27.29  &   27.25  &   4.04  &   5.40  \\
  7352  &  53.1376952  &  --27.7812664  &   31.88  &   28.88  &   26.92  &   26.86  &   26.45  &   26.16  &   6.27  &   4.60  \\
  8033  &  53.1519631  &  --27.7781802  &   99.00  &   31.24  &   28.70  &   26.29  &   26.35  &   25.75  &   7.22  &   6.00  \\
  8301  &  53.1671739  &  --27.7745269  &   35.49  &   29.64  &   27.28  &   27.00  &   27.37  &   27.03  &   4.73  &   4.90  \\
  8664  &  53.1890679  &  --27.7770073  &   99.00  &   28.77  &   26.92  &   26.80  &  \nodata  &  \nodata  &   4.96  &   5.00  \\
  8682  &  53.1887985  &  --27.7770926  &   30.61  &   28.08  &   26.13  &   25.83  &  \nodata  &  \nodata  &   8.01  &   5.00  \\
  8896  &  53.1900057  &  --27.7790583  &   31.79  &   28.44  &   26.94  &   26.90  &  \nodata  &  \nodata  &   4.73  &   5.00  \\
  8961  &  53.1420611  &  --27.7797801  &   99.00  &   30.65  &   28.83  &   26.68  &  \nodata  &  \nodata  &   4.01  &   5.80  \\
  9202  &  53.1383610  &  --27.7786918  &   99.00  &   32.11  &   29.13  &   27.61  &  \nodata  &  \nodata  &   5.41  &   5.70  \\
  9409 $\S$ & 53.1534403 & --27.7661189 & 27.22 & 25.21 & 24.70 & 24.59 & \nodata & \nodata & 5.30 & 3.79 $\ddagger$ \\
  9777  &  53.1702380  &  --27.7628560  &   99.00  &   28.19  &   26.20  &   25.35  &  \nodata  &  \nodata  &   5.04  &   5.40  \\
  9857  &  53.1627716  &  --27.7607679  &   30.87  &   38.76  &   28.58  &   27.12  &  \nodata  &  \nodata  &   3.69  &   5.80  \\
  9983  &  53.1671618  &  --27.7598597  &   30.31  &   27.51  &   25.66  &   25.51  &  \nodata  &  \nodata  &   7.34  &   4.80  \\
 20191  &  53.1725568  &  --27.8137124  &   31.23  &   27.15  &   25.77  &   25.67  &   25.81  &   25.51  &   5.79  &   4.67  \\
 30591  &  53.1553196  &  --27.8151593  &   33.39  &   31.37  &   32.27  &   27.51  &  \nodata  &  \nodata  &   4.99  &   6.70  \\
 32042  &  53.1689736  &  --27.8007244  &   99.00  &   32.72  &   31.75  &   28.84  &  \nodata  &  \nodata  &   4.00  &   5.75  \\
 33003  &  53.1460653  &  --27.7944931  &   99.00  &   99.00  &   31.63  &   28.00  &   27.47  &   27.35  &   4.01  &   6.40  \\
 35506  &  53.1660816  &  --27.7719653  &   30.31  &   32.92  &   31.06  &   27.82  &  \nodata  &  \nodata  &   5.97  &   6.20  \\
 36383  &  53.1677041  &  --27.7681044  &   99.00  &   99.00  &   31.07  &   28.66  &  \nodata  &  \nodata  &   5.58  &   5.80  \\
--101 $\S$ & 53.1770656 & --27.7643556 & 26.23 & 24.52 & 24.20 & 24.22 & \nodata & \nodata & 7.50 & 3.60 $\ddagger$ \\
--102 $\S$ & 53.1431467 & --27.8155017 & 27.95 & 25.10 & 24.17 & 24.14 & \nodata & \nodata & 9.70 & 4.14 $\ddagger$ \\
\enddata

\tablenotetext{\dagger}{Identification from the HUDF catalogs of \citet{beck06}, ACS grism spectra from \citet{malh05}.}
\tablenotetext{*}{Half light radius (R50) measured in \zz-band.} 
\tablenotetext{\S}{Objects from \citet{beck06} \bb-band dropout catalog. Two objects with negative IDs don't have published HUDF IDs.} 
\tablenotetext{\ddagger}{Redshifts from \citet{vanz06}.}



\end{deluxetable}

\clearpage

\clearpage

\clearpage

\clearpage

\clearpage

\end{document}